**Nanocomposite NiO:Pd hydrogen sensors with sub-ppm detection limit and low operating temperature**

M. Kandyla[a,*], C. Chatzimanolis-Moustakas[a], M. Guziewicz[b], and M. Kompitsas[a]

[a]*National Hellenic Research Foundation, Theoretical and Physical Chemistry Institute, 48 Vasileos Constantinou Avenue, 11635 Athens, Greece*

[b]*Institute of Electron Technology, Warsaw 02-668, Poland*

**Abstract**

We present results on the fabrication of chemiresistive nanocomposite NiO:Pd hydrogen sensors, which are able to detect hydrogen concentrations as low as 300 ppb in air, operating at temperatures in the 115 – 145°C range. Thin NiO films were deposited by radio frequency magnetron sputtering on silicon and quartz substrates and their structural and morphological properties were characterized. Pd nanoparticles were added on the NiO surface via pulsed laser deposition and the films were tested as hydrogen sensors before and after Pd deposition. The presence of Pd nanoparticles improved the sensor performance, leading to the detection of hydrogen concentrations at sub-ppm levels. Moreover, the response time of the sensors decreased by a factor between two and four.

* Corresponding author:
M. Kandyla
National Hellenic Research Foundation
Theoretical and Physical Chemistry Institute
48 Vasileos Constantinou Avenue
11635 Athens, Greece
Tel.: +30 210 7273826, Fax: +30 210 7273794
E-mail: kandyla@eie.gr

## 1. Introduction

Hydrogen holds great promise for becoming an alternative fuel for clean energy generation. In this context, hydrogen mass production, distribution, as well as combustion in engines will require advanced control systems to ensure safe storage and usage of the gas, which is colourless and odourless with a lowest explosion limit of 40,000 ppm in air. Therefore, new smart sensors with high sensitivity to hydrogen will be essential. So far many works have been devoted to find suitable materials for hydrogen sensors, in particular metal oxides like $SnO_2$ [1], $WO_3$ [2], ZnO [3], and NiO [4, 5], and also GaN [6]. The successful development of commercial sensors requires high sensitivity, short response time, good repeatability, and reliability.

In this work, we present results on the fabrication of chemiresistive nanocomposite NiO:Pd hydrogen sensors, which are able to detect hydrogen concentrations as low as 300 ppb in air, operating at temperatures in the 115 – 145$^o$C range. NiO is a widely employed non-toxic metal oxide, with excellent chemical stability [7] and ease of fabrication. The use of NiO for gas sensing applications is limited because native NiO is a p-type semiconductor [8], whereas n-type materials are expected to achieve better gas responses than p-type materials [9]. Among the most sensitive pure NiO hydrogen sensors reported so far, thin NiO films, which are able to detect concentrations as low as 500 ppm, operate in the 300 – 650$^o$C temperature range [10, 11]. Pure NiO films have also been shown to operate at temperatures as low as 125$^o$C for higher hydrogen concentrations of 3000 ppm [12]. NiO films with thin Pt overlayers have been employed as chemiresistive hydrogen sensors, detecting concentrations of 500 ppm while operating in the relatively low temperature range of 150 – 420$^o$C, due to the promoting role of the metallic element [13]. Recently, NiO films with embedded Au nanoparticles achieved sensing of hydrogen concentrations as low as 5 ppm, operating in the 100 – 160$^o$C temperature range [14]. Palladium sensors constitute a separate class of hydrogen sensors, which are based on the unique response of palladium to hydrogen [15]. In this work, we combine for the first time palladium with a metal oxide to develop a high-performance hydrogen sensor. By employing high-quality, conductive NiO films with Pd nanoparticles deposited on the surface, we are able to detect hydrogen concentrations at the sub-ppm level for lower operating temperatures than those commonly published in the literature. Consequently, we improve substantially the detection limit of the sensing device, while simultaneously reducing its power consumption. Thus, NiO performance is now comparable to the performance of n-type metal-oxide sensors, making NiO:Pd an attractive material for gas sensing applications.

## 2. Materials and methods

Thin NiO films were deposited by radio frequency magnetron sputtering (RF sputtering) from 3" NiO targets in Ar-O$_2$ plasma, using an RF power of 200 W at room temperature or 300$^o$C. High resistivity Si (001) and quartz wafers were used as substrates. The films were not annealed after deposition. The surface morphology and structural properties of the films were characterized using Scanning Electron Microscopy (SEM) and X-Ray diffraction (XRD). The composition of the films was determined by Energy Dispersive Analysis of X-rays (EDS).

The response of the NiO films to low concentrations of hydrogen was investigated, before and after employing pulsed laser deposition (PLD) to deposit Pd nanoparticles on the surface. For pulsed laser deposition, the NiO films were placed in a vacuum chamber, which was evacuated to a pressure of $10^{-5}$ mbar. A Q-switched Nd:YAG laser system was employed (355 nm wavelength, 10 ns pulse duration), which irradiated a Pd target for 60 s, in order to deposit Pd nanoparticles without covering the NiO surface entirely. During Pd deposition, the NiO films were heated to 100$^{o}$C. The films were not annealed after Pd deposition.

Dynamic sensor response measurements were performed under controlled flows of hydrogen and dry air in the temperature range 115°C - 145°C, in a home-built sensing set-up. During measurements, the NiO samples were heated inside an aluminium chamber and their temperature was continuously monitored using a thermocouple. The electric current through the samples was recorded in real time by a Keithley 485 picoammeter at a bias voltage of 1 V. Changes in current were monitored for various hydrogen concentrations and operating temperatures. The hydrogen concentration in air was calculated from the hydrogen partial pressure in the chamber, measured by an MKS Baratron gauge. In order to achieve low hydrogen concentrations, hydrogen was mixed with dry nitrogen in a premixing chamber, thus achieving dilution factors below $10^{-2}$. The sensor response $S$ was calculated as

$$S = (R_g - R_0)/R_0 \qquad (1),$$

where $R_g$ is the electric resistance of the sample in the presence of hydrogen and $R_0$ is the resistance of the sample in air.

## 3. Results and discussion

The as-deposited NiO films are polycrystalline with *fcc* structure, as indicated by XRD measurements (not shown here). The texture of the films depends on the oxygen content in the Ar-$O_2$ plasma during sputtering. The (111) orientation is observed for deposition in pure Ar or with low oxygen content in the Ar-$O_2$ plasma and the (200) orientation is observed for deposition in pure $O_2$. When a film is deposited in oxygen-rich plasma, then both (111) and (200) orientations appear in the XRD spectrum. For NiO films deposited with oxygen content ranging from 9% to 50% in the Ar-$O_2$ plasma, the lattice parameter increases from 4.194 Å to 4.303 Å, respectively. The composition of the NiO films for the same range of oxygen content in the plasma is calculated from EDS measurements of the film on the silicon substrate and results in an oxygen-to-nickel ratio from 1.1 to 1.23, respectively. Deposition at higher oxygen content in the Ar-$O_2$ plasma does not increase the lattice parameter or the oxygen-to-nickel ratio, but changes the crystal orientation of grains from random to the {200} texture, as well as decreases the lateral size of grains. The attained oxygen-to-nickel ratios result in well-defined p-type conductivity, characteristic of oxygen-rich NiO compounds [5]. In the case of NiO deposition at 300°C, columnar growth in the <111> direction is preferred, and the oxygen-to-nickel ratio is close to 1.1. The electrical properties of NiO films after sputtering and thermal treatment in Ar or $O_2$ were previously studied and reported in Refs. [5, 16].

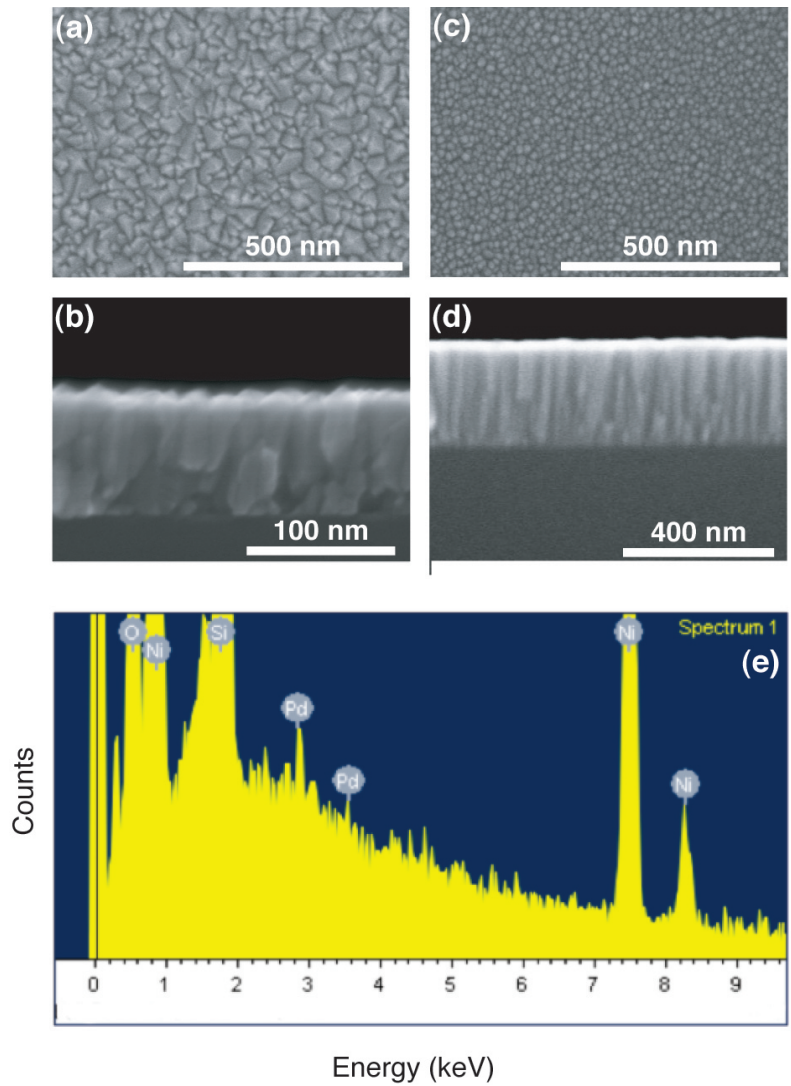

**Figure 1**

Figure 1 shows SEM images of two NiO films, in top view and cross section: (i) a film deposited with 80% $O_2$ content in the Ar-$O_2$ plasma at room temperature on quartz (Figs. 1a and 1b) and (ii) a film deposited with 33% $O_2$ content in the Ar-$O_2$ plasma at 300°C on Si (Figs. 1c and 1d). We observe that deposition at 80% $O_2$ and room temperature results in an average grain size of 33 nm on the film plane (estimated with the Image J software) and a thickness of about 90 nm. On the other hand, deposition at 33% $O_2$ and 300°C results in columnar growth in the (111) direction, an average grain size of 14 nm, and a film thickness of about 300 nm. Similar correlations between NiO film structure and growth temperature were reported in Refs. [17, 18].

The film deposited with 33% $O_2$ content at 300°C performed better for hydrogen sensing. This can be explained by the smaller grain size of this film, which results in an increased surface to volume ratio and therefore a bigger active surface area for sensing [12]. Figure 1e shows an EDS spectrum for this film, after the deposition of Pd nanoparticles on the surface. Only oxygen, nickel, silicon, and palladium peaks appear on the spectrum, indicating the film is free from contaminations. The silicon peak originates from the substrate on which the film is deposited. The atomic concentration of palladium in the sensor film is found to be 0.18%. Similar results were obtained for the palladium concentration of the other NiO:Pd films employed in this study, which was expected since palladium was deposited on all films under identical conditions. Below we present results on hydrogen sensing with the film deposited with 33% $O_2$ content at 300°C.

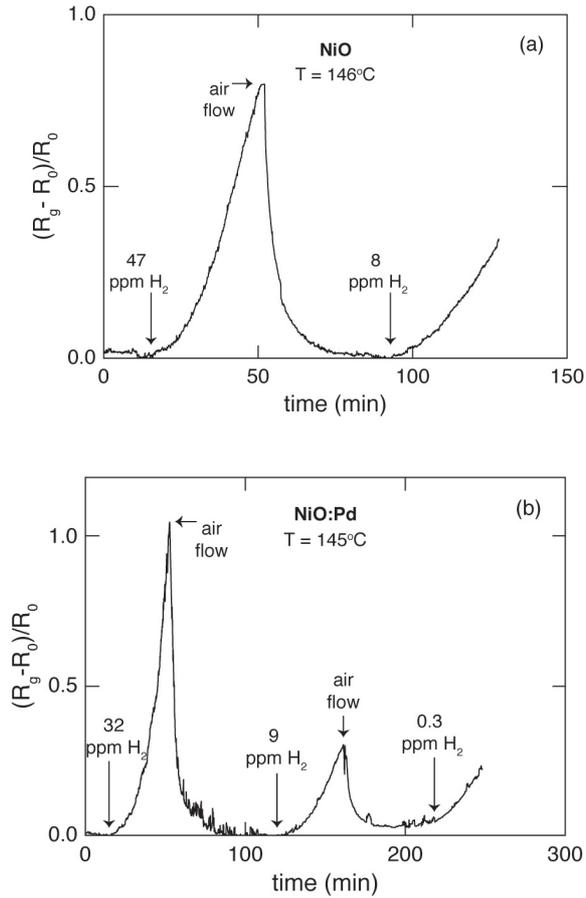

**Figure 2**

The NiO films were tested as hydrogen sensors before and after deposition of Pd nanoparticles on their surface. Because the films conduct electricity as p-type semiconductors, the electric resistance increases during the interaction with hydrogen, which is a reducing gas [19]. This behaviour is demonstrated in Fig. 2a, which shows the response of the as-deposited NiO film to hydrogen. The sensor response is calculated according to Eq. 1. The measurements in Fig. 2a were taken at an operating temperature of 146°C. We observe that the film is able to detect low hydrogen concentrations, as low as 8 ppm, at a relatively low operating temperature with very good signal to noise ratio.

Figure 2b shows the response of the NiO film to hydrogen, after the deposition of Pd nanoparticles on the surface. We observe that the NiO:Pd film is able to detect much lower hydrogen concentrations, as low as 300 ppb, at a similar operating temperature of 145$^{o}$C, compared to the pure NiO film. Moreover, we notice that the purging time needed to recover the active surface is very short. This behavior suggests that the NiO surface plays a dominant role in the electrical properties and the contribution of grain boundaries in conduction is less evident in the measured current.

Figure 3 shows a summary of the maximum response obtained with NiO:Pd films, sputtered with 33% $O_2$ content in the Ar-$O_2$ plasma at 300$^{o}$C, for different operating temperatures and hydrogen concentrations. At 145$^{o}$C, the sensitivity of the films, which is defined as the slope of the sensor response curve versus hydrogen concentration, significantly increases compared with the other two operating temperatures. This suggests that the relevant redox reactions, which constitute the physical mechanism for hydrogen detection with NiO, are activated more efficiently near 145$^{o}$C. These redox reactions consist mainly of exchange of electrons between hydrogen and atmospheric oxygen, which is adsorbed on the surface of NiO [20]. Additionally, the response of the films increases with increasing operating temperature for similar hydrogen concentrations. This is due to the fact that adsorption-desorption kinetics, which affect the sensor performance, depend on the operating temperature [21, 22] and apparently become more efficient near 145°C. For the operating temperature of 115°C, the hydrogen detection limit is 6.5 ppm. This is one of the lowest operating temperatures found in the literature for NiO sensors, which typically operate between 300 – 650°C, as mentioned in the Introduction. For the operating temperatures of 132°C and 145°C, the hydrogen detection limit is 700 and 300 ppb, respectively.

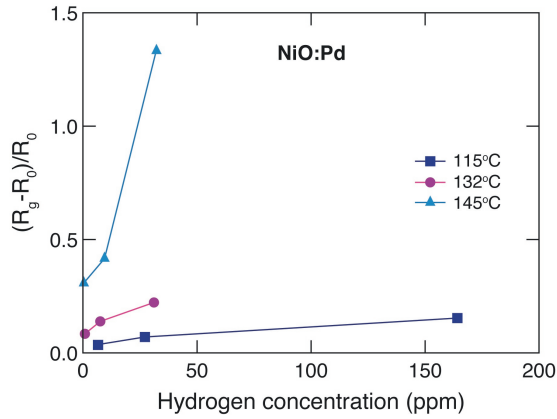

**Figure 3**

Table 1 shows hydrogen sensing results for the NiO film deposited with 33% $O_2$ content in the Ar-$O_2$ plasma at 300°C on quartz, before and after Pd nanoparticle deposition, where the sensor response time is also indicated. The response time is defined as the time interval between 10% and 90% of the total signal change. We observe that the presence of Pd nanoparticles on the surface reduces the response time of the sensor by a factor between two and four. The improvement in the response time is more pronounced for the lower temperature of 130°C, making the NiO:Pd sensor operation more favorable for lower operating temperatures, compared to pure NiO films. This result promotes the development of low-power gas sensors.

Concerning the repeatability of the measurements for low hydrogen concentrations, all samples presented in this work were able to detect 700 ppb of hydrogen at the operating temperature of 130$^o$C, after the deposition of Pd nanoparticles on their surface. Table 2 presents the maximum response of the different samples at 700 ppb of hydrogen at T ≈ 130$^o$C, which is calculated according to Eq. 1. We observe that the film deposited with 80% $O_2$ content in the Ar-$O_2$ plasma performs lesser than the other two, as was mentioned earlier. In all cases the concentration of 700 ppb was clearly detected. Due to experimental constraints of gas flow control, measurements for the 300 ppb hydrogen concentration were possible only for one sample, therefore we are not able to make a direct comparison of the response of different films for this concentration. However, the film deposited with 80% $O_2$ content in the Ar-$O_2$ plasma, was able to clearly detect 500 ppb of hydrogen at 144$^o$C.

The partial coverage of the NiO surface with Pd nanoparticles decreases the hydrogen detection limit to sub-ppm levels and also decreases significantly the sensor response time. The nanoparticles act as catalysts for the reaction between hydrogen and atmospheric oxygen adsorbed on the NiO surface, which results in the change of the electric properties of the film in the presence of hydrogen. Metallic nanoparticles dissociate $O_2$ during the initial adsorption process and $H_2$ during the sensing process, resulting in highly activated atomic oxygen and hydrogen species, which react more effectively [23]. Additionally, atmospheric oxygen gets adsorbed on the surface of the metallic nanoparticles and reacts with hydrogen. Therefore, the presence of metallic nanoparticles increases the active surface area of the sensor. Comparing the performance of Pd nanoparticles with Au nanoparticles presented in previous works [14, 24], we conclude that Pd acts as a better catalyst for the performance of nanocomposite NiO hydrogen sensors.

## 4. Conclusions

Thin NiO films were deposited by RF magnetron sputtering and employed as resistive hydrogen sensors. Hydrogen concentrations of a few ppm in air were detected at operating temperatures ranging from 130°C to 145°C. Subsequently, Pd nanoparticles were deposited on the NiO surface by pulsed laser deposition. The presence of Pd nanoparticles improved the sensor performance, leading to the detection of hydrogen concentrations at the sub-ppm level, as low as 300 ppb. The response time of the sensors decreased by a factor between two and four. The NiO:Pd sensors were able to operate at 115°C, which is one of the lowest reported operating temperatures for NiO-based hydrogen sensors. This reduces significantly the power consumption of the devices, leading to the development of portable sensors, and provides additional safety during the operation of the sensor in the ambient of a flammable and explosive (at elevated temperatures) gas, as hydrogen. According to our knowledge, the combination of low operating temperatures and ppb-level measurable hydrogen concentrations presented in this work, is one of the best found in the literature.

**Acknowledgements**

M.G. acknowledges financial support from the EU European Regional Development Fund, through the grant Innovative Economy POIG.01.03.01-00-159/08 (InTechFun). The authors would like to thank E. Chatzitheodoridis for EDS measurements.

# References

[1] Barsan N, Koziej D, Weimar U. Metal oxide-based gas sensor research: How to?. Sensor Actuat B 2007; 121:18–35.

[2] Ippolito SJ, Kandasamy S, Kalantar-zadeh K, Wlodarski W. Hydrogen sensing characteristics of $WO_3$ thin film conductometric sensors activated by Pt and Au catalysts. Sensor Actuat B 2005; 108:154–8.

[3] Pandis Ch, Brilis N, Bourithis E, Tsamakis D, Ali H, Krishnamoorthy S, *et al.* Low–temperature hydrogen sensors based on Au nanoclusters and Schottky contacts on ZnO films deposited by pulsed laser deposition on Si and $SiO_2$ substrates. IEEE Sens J 2007; 7:448–54.

[4] Brilis N, Foukaraki C, Bourithis E, Tsamakis D, Giannoudakos A, Kompitsas M, *et al.* Development of NiO-based thin film structures as efficient $H_2$ gas sensors operating at room temperatures. Thin Solid Films 2007; 515:8484–9.

[5] Guziewicz M, Grochowski J, Borysiewicz M, Kamińska E, Domagała JZ, Rzodkieiwcz W, *et al.* Electrical and optical properties of NiO films deposited by magnetron sputtering. Opt Appl 2011; 41:431-40.

[6] Kouche AEL, Lin J, Law ME, Kim S, Kim BS, Ren F, *et al.* Remote sensing system for hydrogen using GaN Schottky diodes. Sensor Actuat B 2005; 105:329–33.

[7] Kumagai H, Matsumoto M, Toyoda K, Obara M. Preparation and characteristics of nickel oxide thin film by controlled growth with sequential surface chemical reactions. J Mater Sci Lett 1996; 15:1081-3.

[8] Adler D, Feinleib J. Electrical and optical properties of narrow-band materials. Phys Rev B 1970; 2;3112-34.

[9] Barsan N, Simion C, Heine T, Pokhrel S, Weimar U. Modeling of sensing and transduction for p-type semiconducting metal oxide based gas sensors. J Electroceram 2010; 25:11-9.

[10] Gu H, Wang Z, Hu Y. Hydrogen gas sensors based on semiconductor oxide nanostructures. Sensors 2012; 12:5517-50.

[11] Steinebach H, Kannan S, Rieth L, Solzbacher F. $H_2$ gas sensor performance of NiO at high temperatures in gas mixtures. Sensor Actuat B 2010; 151:162-8.

[12] Soleimanpour AM, Hou Y, Jayatissa AH. Evolution of hydrogen gas sensing properties of sol-gel derived nickel oxide thin film. Sensor Actuat B 2013; 182:125-33.

[13] Hotovy I, Huran J, Siciliano P, Capone S, Spiess L, Rehacek V. Enhancement of $H_2$ sensing properties of NiO-based thin films with a Pt surface modification. Sensor Actuat B 2004; 103:300-11.

[14] Fasaki I, Kandyla M, Tsoutsouva MG, Kompitsas M. Optimized hydrogen sensing properties of nanocomposite NiO:Au thin films grown by dual pulsed laser deposition. Sensor Actuat B 2013; 176:103-9.

[15] Soundarrajan P, Schweighardt F. Hydrogen sensing and detection. In: Gupta RB, editor. Hydrogen fuel, production, transport, and storage, CRC Press; 2008, p. 495–534.

[16] Guziewicz M, Jung W, Grochowski J, Borysiewicz M, Gołaszewska K, Kruszka R, *et al.* Influence of thermal and gamma radiation on electrical properties of thin NiO films formed by RF sputtering. Procedia Engineering 2011; 25:367–70.

[17] Karpinski A, Ferrec A, Richard-Plouet M, Cattin L, Djouadi MA, Brohan L, *et al.* Deposition of nickel oxide by direct current reactive sputtering: Effect of oxygen partial pressure. Thin Solid Films 2012; 520:3609-13.

[18] Chen HL, Lu YM, Hwang WS. Characterization of sputtered NiO thin films. Surf Coat Tech 2005; 198:138–42.

[19] Morisson SR. Selectivity in semiconductor gas sensors. Sensor Actuat 1987; 12:425-40.

[20] Fasaki I, Kandyla M, Kompitsas M. Properties of pulsed laser deposited nanocomposite NiO:Au thin films for gas sensing applications. Appl Phys A 2012; 107:899 – 904.

[21] Ahn M-W, Park K-S, Heo J-H, Park JG, Kim DW, Choi KJ, *et al.* Gas sensing properties of defect-controlled ZnO-nanowire gas sensor. Appl Phys Lett 2008; 93:263103.

[22] Lupan O, Ursaki VV, Chai G, Chow L, Emelchenko GA, Tiginyanu IM, *et al.* Selective hydrogen gas nanosensor using individual ZnO nanowire with fast response at room temperature. Sensor Actuat B 2010; 144:56-66.

[23] Korotcenkov T, Gulina LB, Cho BK, Han SH, Tolstoy VP. $SnO_2$-Au nanocomposite synthesized by successive ionic layer deposition method: Characterization and application in gas sensors. Mater Chem Phys 2011; 128:433-41.

[24] Gaspera ED, Guglielmi M, Martucci A, Giancaterini L, Cantalini C. Enhanced optical and electrical gas sensing response of sol-gel based NiO-Au and ZnO-Au nanostructured thin films. Sensor Actuat B 2012; 164:54-63.

**Figure captions**

Figure 1. Top view and cross section SEM images of NiO films deposited by RF sputtering with RF power of 200 W for (a, b) 80% $O_2$ in the Ar-$O_2$ plasma at room temperature on quartz and for (c, d) 33% $O_2$ in the Ar-$O_2$ plasma at 300$^o$C on Si. (e) EDS spectrum of the NiO film shown in Figs. 1c and 1d, after Pd deposition on the surface.

Figure 2. Hydrogen sensing in air under dynamic flow conditions obtained with a NiO film sputtered with 33% $O_2$ content in the Ar-$O_2$ plasma at 300$^o$C on Si, (a) before and (b) after Pd nanoparticle deposition.

Figure 3. Summary of sensing results for different operating temperatures and hydrogen concentrations in air, obtained for NiO films sputtered with 33% $O_2$ content in the Ar-$O_2$ plasma at 300$^o$C on Si, with Pd nanoparticles.

**Table captions**

Table 1: Hydrogen detection limit and response times of NiO films before and after Pd deposition, for various operating temperatures.

Table 2: Response of different NiO:Pd samples to 700 ppb of hydrogen in air at the operating temperature T ≈ 130°C.

**Table 1**

| Sample | O$_2$ content in plasma (%) | Pd nanoparticles | Detection limit (ppm) | Response time (min) | Operating temperature (°C) |
|---|---|---|---|---|---|
| NiO | 33 | No | 5 | 46 | 130 |
|  |  | No | 4 | 26 | 143 |
|  |  | Yes | 0.7 | 10 | 130 |
|  |  | Yes | 0.7 | 11 | 145 |

**Table 2**

| Substrate | O$_2$ content in plasma (%) | Sample | Operating temperature (°C) | Hydrogen concentration (ppm) | Response |
|---|---|---|---|---|---|
| Quartz | 33 | NiO:Pd | 130 | 0.7 | 0.135 |
| Silicon | 33 | NiO:Pd | 132 | 0.7 | 0.084 |
| Quartz | 80 | NiO:Pd | 130 | 0.7 | 0.064 |